\documentclass[twocolumn,english,aps,prd,showpacs,superscriptaddress]{revtex4-1}
\usepackage[T1]{fontenc}
\usepackage[latin9]{inputenc}
\usepackage{xcolor}
\usepackage{amsmath}
\usepackage{amssymb}
\usepackage{graphicx}
\PassOptionsToPackage{normalem}{ulem}
\usepackage{ulem}

\makeatletter

\providecolor{lyxadded}{rgb}{0,0,1}
\providecolor{lyxdeleted}{rgb}{1,0,0}

\usepackage{babel}

\makeatother

\usepackage{babel}
\begin{document}

\title{RKKY interaction in three-dimensional electron gases with linear spin-orbit coupling}

\author{Shi-Xiong Wang}

\affiliation{Department of Physics, Institute of Solid State Physics, and Center
for Computational Sciences, Sichuan Normal University, Chengdu, Sichuan 610066, China}

\author{Hao-Ran Chang}
\email{hrchang@mail.ustc.edu.cn}
\affiliation{Department of Physics, Institute of Solid State Physics, and Center
for Computational Sciences, Sichuan Normal University, Chengdu, Sichuan 610066, China}
\affiliation{Department of Physics, McGill University, Montreal, Quebec H3A 2T8, Canada}

\author{Jianhui Zhou}
\email{jianhuizhou1@gmail.com}
\affiliation{Department of Physics, The University of Hong Kong, Pokfulam Road, Hong Kong, China}

\begin{abstract}
We theoretically study the impacts of linear spin-orbit coupling (SOC)
on the Ruderman-Kittel-Kasuya-Yosida interaction between magnetic
impurities in two kinds of three-dimensional noncentrosymmetric systems.
It has been found that linear SOCs lead to the Dzyaloshinskii-Moriya interaction and the Ising interaction, in
addition to the conventional Heisenberg interaction. These interactions
possess distinct range functions from three-dimensional electron gases
and Dirac/Weyl semimetals. In the weak SOC limit, the Heisenberg
interaction dominates over the other two interactions in a moderately
large region of parameters. Sufficiently strong Rashba SOC makes the
Dzyaloshinskii-Moriya interaction or the Ising interaction dominate
over the Heisenberg interaction in some regions. The change in topology of the Fermi
surface leads to some quantitative changes in periods of oscillations
of range functions. The anisotropy of Ruderman-Kittel-Kasuya-Yosida
interaction in bismuth tellurohalides family BiTe$X$ ($X$ = Br,
Cl, and I) originates from both the specific form of Rashba SOC and
the anisotropic effective mass. Our work provides some insights into
understanding observed spin textures and the application of these materials in spintronics. 
\end{abstract}

\date{\today}
\maketitle

\section{introduction}

Inversion symmetry may be broken at the surface or interface, and
the resulting electric fields entangle momentum of an electron to
its spin, leading to the Rashba spin-orbit coupling (SOC) \cite{Bychkov1984jpc}.
The Rashba SOC has mostly been studied in two-dimensional electron
or hole gases in semiconductor heterostructures \cite{winkler2003SOC},
the surface states of conventional metals \cite{LaShell1996PRL},
and even the two-dimensional electron gases (2DEG) at the interface or surfaces
of complex oxides \cite{Ohtomo2004Nature,Caviglia2010PRL,BenShalom2010PRL,king2014NC,Khalsa2013PRB,Zhou2015PRB}.
The Rashba SOC, $H_{R}=\alpha_{R}\left(\boldsymbol{\sigma}\times\boldsymbol{k}\right)\cdot\hat{z}$,
plays a central role in a vast number of phenomena in condensed matter
physics, such as spin Hall effect, spin galvanic effect, spin\textendash orbit
torques, chiral magnetic textures, topological phases of matter, and
Majorana fermions \cite{Fert2013NNano,Manchon2015NM,Sinova2015RMP,soumyanarayanan2016Nature}.
In particular, the interplay between SOC and magnetism is of increasing
importance. Rashba SOC could induce a chiral Dzyaloshinskii\textendash Moriya
(DM) interaction \cite{DZYALOSHINSKY1958jpcs,Moriya1960PR}, which
leads to spin helix, chiral domain walls, and magnetic skyrmions.
In addition, Ruderman-Kittel-Kasuya-Yosida (RKKY) interaction, an indirect exchange interaction between two magnetic impurities mediated by conduction electrons, 
may provide an alternative mechanism for the DM interaction.  
The strength of RKKY interaction, characterized by the range function, usually oscillates as a function of distance between magnetic impurities 
with the period determined by the Fermi level, the effective mass of conduction electrons, and other material-dependent parameters~\cite{Ruderman1954Kittel,Kasuya1956,Yosida1957}. 

Recently, the Rashba SOC in three-dimensional (3D) systems lacking inversion
symmetry has been receiving significant interest~\cite{Liebmann2016AM,Elmers2016PRB}. In the bismuth tellurohalides
family BiTe$X$ ($X$ = Br, Cl, and I), the structure inversion asymmetry
results from the asymmetric stacking of Bi, Te, and $X$ layers, in
which an exceptionally large Rashba parameter had been revealed by
angle-resolved photoemission spectroscopy measurements \cite{Ishizaka2011BiTeI,Sakano2013PRL}.
For example, the strength of Rashba SOC reaches the value of $3.9-4.3~\:\mathrm{eV\mathring{A}}$
for BiTeI as large as that of the surface state of topological insulators
\cite{Sakano2013PRL}. Symmetry analysis suggests that the conduction
electrons in $B_{20}$ compounds (a class of cubic helimagnets)
\cite{Kang2015PRBb20} and noncentrosymmetric metal $\mathrm{Li}_{2}\left(\mathrm{Pd}_{1-x},\mathrm{Pt}_{x}\right)_{3}\mathrm{B}$
\cite{Mineev2010PRB} possess a strong SOC of the form $\boldsymbol{k}\cdot\boldsymbol{\sigma}$.
Many nontrivial spin textures have already been observed in BiTe$X$
\cite{Ishizaka2011BiTeI,Landolt2012PRL,Maabeta2016NC} as well as
$B_{20}$ compounds \cite{Uchida2006Science,Roszler2006Nature,Muhlbauer2009Science,Yu2010Nature}.
To thoroughly understand the observed various interesting spin textures
and to engineer exotic states of quantum matter, we would like to systematically study 
the effects of linear SOCs on the RKKY interaction. 

In this paper, we find that 3D linear SOCs give rise to the DM interaction, the Ising interaction, and
the Heisenberg interaction, which possess entirely different range
functions from those of 3D electron gases (3DEG) and Dirac/Weyl
semimetals. In the weak SOC limit, the Heisenberg interaction dominates
over the other two interactions for a relatively large region of parameters.
As the Rashba SOC increases, sufficiently strong Rashba SOC makes
the DM interaction or the Ising interaction more favorable than the
Heisenberg interaction. As the topology of the Fermi surface changes,
the periods of oscillations of range functions vary accordingly.
In the family BiTe$X$, both the Rashba SOC and the anisotropic
effective mass contribute to the anisotropy of the RKKY interaction. 

The rest of this paper is organized as follows. In Sec. \ref{sec:rkkysoc},
we outline the formalism for the RKKY interaction. In Sec. \ref{sec:ncs},
we derive the exact analytical expressions of RKKY interactions for
noncentrosymetric metals and the approximate expressions in weak SOC
limit and in long-range limit. In Sec. \ref{sec:3drashba}, we reveal
the general features of RKKY interactions for 3D Rashba semiconductors numerically and find the analytical expressions 
in weak Rashba SOC limit. In Sec. \ref{sec:conclus}, the main results of this paper
are summarized. Finally, in the appendices, we give the detailed calculations
of range functions of the RKKY interaction.  

\section{\label{sec:rkkysoc} rkky interaction and spin-orbit coupling}

The RKKY interaction is an indirect exchange interaction between two
localized spins via the spin polarization of conduction electrons
~\cite{Ruderman1954Kittel,Kasuya1956,Yosida1957}. In recent years,
it has been shown that the RKKY interaction plays an important role
in giant magnetoresistance in multilayer structures \cite{Parkin1990PRL},
ferromagnetism in diluted magnetic semiconductors \cite{Jungwirth2006RMP},
topological phases, and Majorana fermions \cite{Manchon2015NM}. We
restrict ourselves to a pair of magnetic impurities, $\boldsymbol{S}_{1}$
and $\boldsymbol{S}_{2}$ located at $\boldsymbol{R}_{1}$ and $\boldsymbol{R}_{2}$,
respectively. We assume the interaction between magnetic impurities
and conduction electrons is described by the standard $s$-$d$ interaction~\cite{sdInt}:
\begin{equation}
H_{s-d}=J\sum_{l=1,2}\boldsymbol{S}_{l}\cdot\boldsymbol{\sigma}\delta\left(\boldsymbol{r}-\boldsymbol{R}_{l}\right),\label{sdInts}
\end{equation}
where $\boldsymbol{\sigma}=\left(\sigma_{x},\sigma_{y},\sigma_{z}\right)$
denotes for the vector of Pauli spin matrices and $J$ refers to the
strength of the $s$-$d$ interaction. At zero temperature the indirect
exchange interaction between these two localized spins mediated by
itinerant electrons is given by \cite{ChangHR2015PRB} 
\begin{align}
H_{\mathrm{RKKY}} & =-\frac{J^{2}}{\pi}\mathrm{Im}\int_{-\infty}^{\varepsilon_{F}}\mathrm{d}\varepsilon\mathrm{Tr}\bigl[\left(\boldsymbol{S}_{1}\cdot\boldsymbol{\sigma}\right)G\left(\boldsymbol{R};\varepsilon+i0^{+}\right)\nonumber \\
 & \times\left(\boldsymbol{S}_{2}\cdot\boldsymbol{\sigma}\right)G\left(-\boldsymbol{R};\varepsilon+i0^{+}\right)\bigr],\label{eq:rkky}
\end{align}
where $\boldsymbol{R}=\boldsymbol{R}_{2}-\boldsymbol{R}_{1}$, $\varepsilon_{F}$
is the Fermi energy, and Tr represents the trace over the spin degree
of freedom of itinerant electrons. The Green's functions in the energy-coordinate
representation can be obtained from the Fourier transform: 
\begin{equation}
G\left(\pm\boldsymbol{R};\varepsilon+i0^{+}\right)=\int\frac{\mathrm{d}^{3}\boldsymbol{k}}{\left(2\pi\right)^{3}}G\left(\boldsymbol{k};\varepsilon+i0^{+}\right)\exp\left(\pm i\boldsymbol{k}\cdot\boldsymbol{R}\right),
\end{equation}
where $G^{-1}\left(\boldsymbol{k};\varepsilon+i0^{+}\right)=\varepsilon+i0^{+}-H_{0}\left(\boldsymbol{k}\right)$
is the inverse of momentum-space Green's function, $H_{0}\left(\boldsymbol{k}\right)$
is the Hamiltonian for the noninteracting conduction electrons, and
$0^{+}$ is a positive infinitesimal. 

\begin{figure}
\includegraphics[scale=0.45]{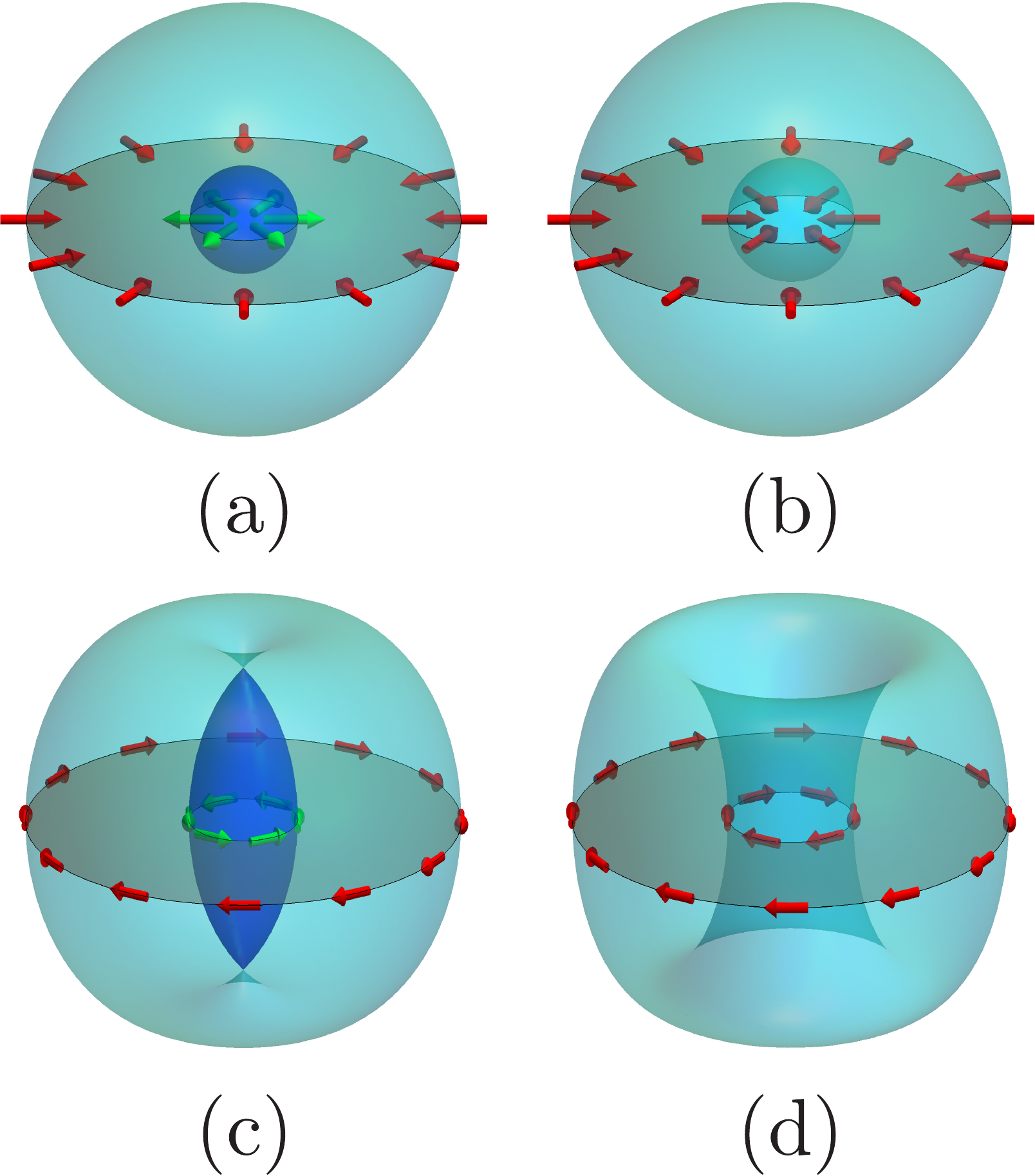}\caption{(Color online) Distinct spin textures at Fermi surfaces of 3D noncentrosymmetric
metals (a) and (b) and 3D Rashba semiconductors (c) and (d).
For (a) and (c), the Fermi energy is positive $\varepsilon_{F}>0$, while for (b) and (d), $\varepsilon_{F}<0$.  
The green and red arrows show the spin directions at the Fermi energy for a given
$k_{z}$ as a function of the two momenta, $k_{x}$ and $k_{y}$.
The topology of Fermi surface changes at $\varepsilon_{F}=0$ from two Fermi surfaces with different spin directions in (a) and (c) to a closed Fermi surface in (b) and (d) or vice versa. \label{Fig1}}
\end{figure}

The specific expressions of the RKKY interactions strongly depend
on the spatial dimensionality of systems \cite{Aristov1997PRB,Litvinov1998PRB},
energy dispersion of the conduction carriers, and the property of
Rashba SOC. For instance, the RKKY interaction for 3D
Dirac electrons with linear dispersion in the context of Dirac/Weyl
semimetals \cite{ChangHR2015PRB,Hosseini2015PRBWeyl,Mastrogiuseppe2016PRB}
greatly differs from that of the conventional 3DEG \cite{Ruderman1954Kittel,Kasuya1956,Yosida1957}. The
influence of the Rashba SOC had been extensively examined in both one-dimensional electron gases
(1DEG)~\cite{Imamura2004PRB,Schulz2009PRB,ZhuJJ2010PRB,Klinovaja2013PRB}
and 2DEG \cite{Imamura2004PRB,Lyu2007JAP,Simon2008PRB,LaiHH2009PRB,Chesi2010PRB,Kernreiter2013PRL}.
In the present paper, we shall focus on the impacts of linear SOC
on the RKKY interactions in both 3D noncentrosymmetric metals and
Rashba semiconductors BiTe$X$. In the former, each component of spin
is parallel to momentum and forms hedgehog spin texture, as shown
in Figs.~$\ref{Fig1}$(a) and $\ref{Fig1}$(b). In the latter, for a given $k_{z}$, the
in-plane components of spin and momenta are perpendicular to each
other as the manner of 2DEG with linear Rashba SOC [see Figs.~$\ref{Fig1}$(c) and $\ref{Fig1}$(d)].
Note that the topology of the Fermi surface changes at $\varepsilon_{F}=0$.
 
\section{\label{sec:ncs} three-dimensional noncentrosymmetric metals}

\begin{figure}
\includegraphics[scale=0.5]{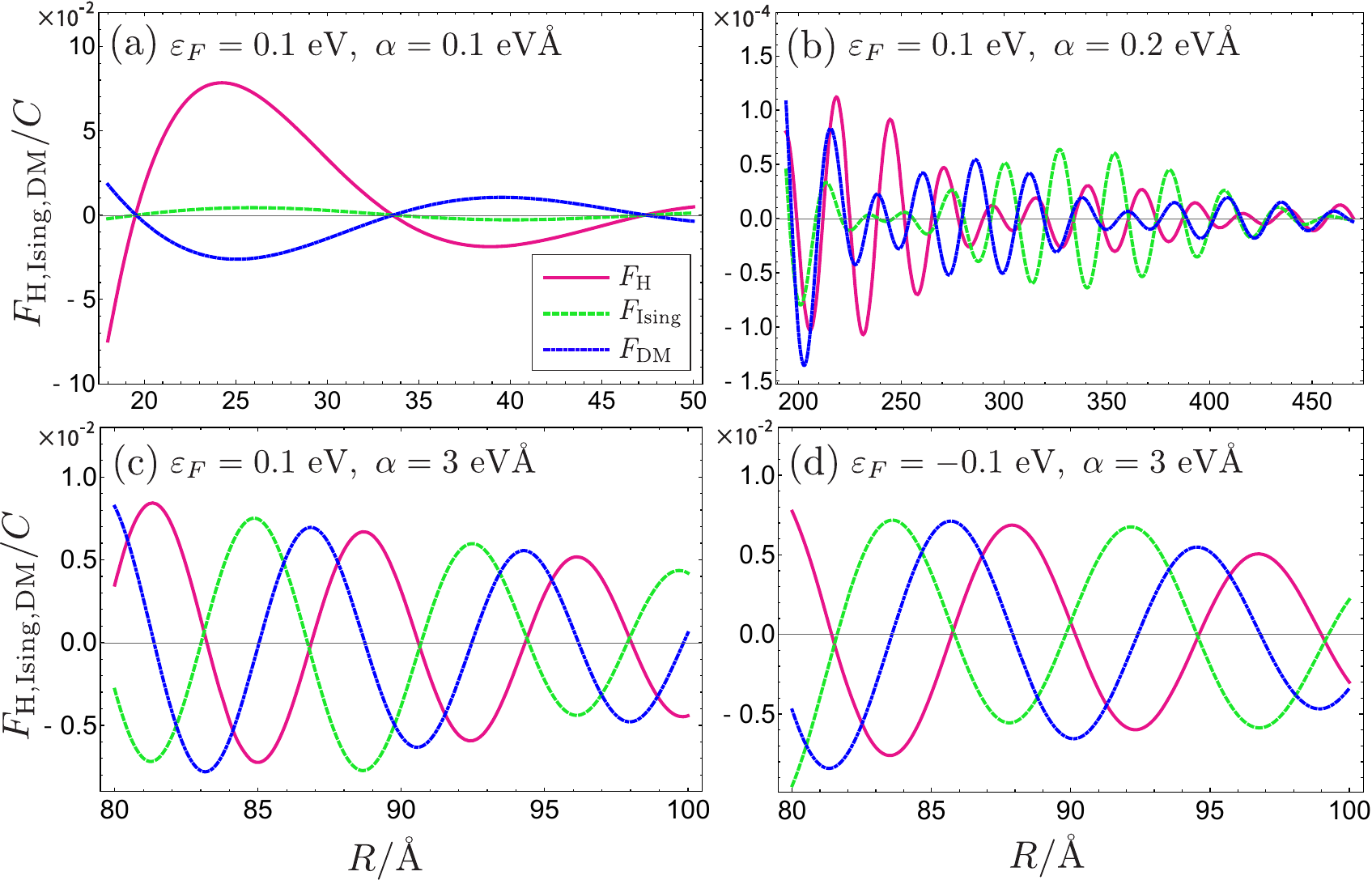}
\caption{The exact range functions of RKKY interactions in 3D noncentrosymmetric
metals for different values of SOCs. The change in topology of the
Fermi surface leads to quantitative changes in the periods of oscillations
(c) and (d). The effective mass is equal to half of the electron rest mass.
\label{Fig2} }
\end{figure}

Three-dimensional noncentrosymmetric metals, such as $B_{20}$ compounds
\cite{Kang2015PRBb20,cubicRashba} and the $\mathrm{Li}_{2}\left(\mathrm{Pd}_{1-x},\mathrm{Pt}_{x}\right)_{3}\mathrm{B}$
family \cite{Samokhin2008PRB}, can be effectively described by the
following Hamiltonian 
\begin{equation}
H_{0}^{\mathrm{NCS}}=\frac{k^{2}}{2m}+\alpha\boldsymbol{k}\cdot\boldsymbol{\sigma},\label{eq:model1}
\end{equation}
and the energy dispersion reads
\begin{equation}
\varepsilon_{\lambda\boldsymbol{k}}=\frac{k^{2}}{2m}+\lambda\alpha k,
\end{equation}
where $m$ is the effective mass, $\boldsymbol{k}=\left(k_{x},k_{y},k_{z}\right)$,
and $k=\sqrt{k_{x}^{2}+k_{y}^{2}+k_{z}^{2}}$. $\lambda=\pm1$ denotes
for the two energy bands with opposite chirality, $\alpha>0$ represents
the strength of SOC. The superscript $\mathrm{NCS}$ stands for the
noncentrosymmetric metals. For a positive Fermi energy $\varepsilon_{F}>0$,
there are two distinct Fermi wave vectors $k_{F\pm}=\mp m\alpha+\sqrt{m^{2}\alpha^{2}+2m\varepsilon_{F}}$,
characterizing the two Fermi surfaces in Fig.~$\ref{Fig1}$(a). On
the other hand, for $-m\alpha^{2}/2<\varepsilon_{F}<0$, only one
closed Fermi surface exists in Fig.~$\ref{Fig1}$(b), but is characterized
by two Fermi wave vectors $k_{F\pm}=m\alpha\pm\sqrt{m^{2}\alpha^{2}+2m\varepsilon_{F}}>0$.
As will be shown below, the change in topology of the Fermi surface
leads to some quantitative modifications to the RKKY interactions
[see Figs.~$\ref{Fig2}$(c) and $\ref{Fig2}$(d)]. 

After some straightforward calculations, one gets the corresponding
Berry curvatures of Bloch electrons as 
\begin{equation}
\boldsymbol{\Omega}_{\pm}\left(\boldsymbol{k}\right)=\mp\frac{\boldsymbol{k}}{2k^{3}},
\end{equation}
which can be viewed as effective magnetic monopoles in the momentum
space, similar to those in 3D Weyl semimetals~\cite{XiaoRMPBerry2010}. 
One thus expects that the electrons in noncentrosymmetric metals inherit the properties of electrons from both the conventional
3DEG and the Dirac/Weyl semimetals \cite{Wan2011PRB}. 

According to Eq.~$\left(\ref{eq:model1}\right)$, one directly evaluates
the Green's function in the momentum space as 
\begin{equation}
G\left(\boldsymbol{k};\varepsilon+\mathrm{i}0^{+}\right)=\frac{1}{2}\sum_{\lambda=\pm}\frac{\sigma_{0}+\lambda\hat{\boldsymbol{k}}\cdot\boldsymbol{\sigma}}{\varepsilon-\varepsilon_{\lambda\boldsymbol{k}}+i0^{+}},\label{eq:GFm-1}
\end{equation}
where $\hat{\boldsymbol{k}}=\boldsymbol{k}/k$ denotes for the unit
vector parallel to $\boldsymbol{k}$ and $\sigma_{0}$ refers to the
identity matrix. Due to the rotational symmetry of itinerant electrons
described by Eq.~$\left(\ref{eq:model1}\right)$, without loss of
generality, we align the two magnetic impurities with the $j$ axis,
i.e., $\boldsymbol{R}=R\boldsymbol{e}_{j}$, with $R$ being the distance
between two magnetic impurities. After carrying out Fourier transform,
we decompose the real-space Green's function into two parts as
\begin{align}
G\left(\pm\boldsymbol{R};\varepsilon+i0^{+}\right) & =G_{0}\sigma_{0}\pm G_{R}\sigma_{j},\label{eq:GFr-1}
\end{align}
where the functions are given as
\begin{align}
G_{0} & =\frac{-m\exp\left(i\xi\right)}{2\pi R\left(\xi+\mathrm{i}0^{+}\right)}\left(\xi\cos\zeta+i\zeta\sin\zeta\right),\nonumber \\
G_{R} & =\frac{m\exp\left(i\xi\right)}{2\pi R\left(\xi+i0^{+}\right)}\left[\left(i\xi-1\right)\sin\zeta+\zeta\cos\zeta\right],
\end{align}
with two dimensionless parameters $\zeta=m\alpha R$ and $\xi=\sqrt{m^{2}\alpha^{2}+2m\varepsilon}R$.
The second term $G_{R}\sigma_{j}$ in Eq.~$\left(\ref{eq:GFr-1}\right)$
comes from the Rashba SOC and greatly modifies the RKKY interactions.
Consequently, the RKKY interaction contains three terms: Heisenberg
interaction, Ising interaction, and DM interaction:
\begin{equation}
H_{\mathrm{RKKY}}^{\mathrm{NCS}}=F_{\mathrm{H}}\boldsymbol{S}_{1}\cdot\boldsymbol{S}_{2}+F_{\mathrm{Ising}}S_{1}^{j}S_{2}^{j}+F_{\mathrm{DM}}\left(\boldsymbol{S}_{1}\times\boldsymbol{S}_{2}\right)_{j},\label{eq:ncsrkky}
\end{equation}
where these three range functions are given by
\begin{align}
F_{\mathrm{H}} & =-\frac{2J^{2}}{\pi}\mathrm{Im}\int_{-\infty}^{\varepsilon_{F}}\left(G_{0}^{2}+G_{R}^{2}\right)\mathrm{d}\varepsilon,\nonumber \\
F_{\mathrm{Ising}} & =\frac{4J^{2}}{\pi}\mathrm{Im}\int_{-\infty}^{\varepsilon_{F}}G_{R}^{2}\mathrm{d}\varepsilon,\nonumber \\
F_{\mathrm{DM}} & =\frac{4J^{2}}{\pi}\mathrm{Im}\int_{-\infty}^{\varepsilon_{F}}iG_{0}G_{R}\mathrm{d}\varepsilon.\label{FhIdm}
\end{align}
The Rashba SOC has two obvious impacts on the RKKY interaction. First,
it modifies the Heisenberg interaction in Eq.~$\left(\ref{FhIdm}\right)$.
Second, both the Ising interaction and the DM interaction entirely
originate from the Rashba SOC. It should be emphasized that the DM
vector along $\boldsymbol{R}=R\boldsymbol{e}_{j}$ is compatible with
the symmetry of $B_{20}$ compounds \cite{Kang2015PRBb20}. Therefore,
the RKKY interaction here may provide us an alternative physical origin
for the DM interaction, which is essential for the chiral spin textures
\cite{Uchida2006Science,Roszler2006Nature,Muhlbauer2009Science,Yu2010Nature}.
Physically, the cross-product nature of the DM interaction implies that an inversion operation about the center of the joint line 
would exchange two neighboring spins such that the DM interaction flips sign. The DM interaction would disappear if the two localized
spins are parallel or antiparallel. However, the Heisenberg interaction
and Ising interaction tend to align neighboring spins and are unchanged
under this operation, respecting the inversion symmetry. 
After performing these complex integrals, one has the exact analytical
range functions in Eq.~$\left(\ref{FhIdm}\right)$ (the detailed calculations can be found in Appendix \ref{sec:AppNCS}):
\begin{align}
 & F_{\mathrm{H}}\left(\xi_{F},\zeta\right)\nonumber \\
= & \frac{C}{\left(k_{F}R\right)^{4}}\left\{ 2\xi_{F}\cos\left(2\zeta\right)\cos\left(2\xi_{F}\right)\right.\nonumber \\
 & -\left[4\zeta\sin\left(2\zeta\right)+3\cos\left(2\zeta\right)-2\right]\sin\left(2\xi_{F}\right)\nonumber \\
 & -2\left.\left[2\zeta^{2}\cos\left(2\zeta\right)+g\left(\zeta\right)\right]\mathrm{si}\left(2\xi_{\mathrm{F}}\right)\right\} ,\nonumber \\
 & F_{\mathrm{\mathrm{Ising}}}\left(\xi_{F},\zeta\right)\nonumber \\
= & \frac{C}{\left(k_{F}R\right)^{4}}\left\{ 2\xi_{F}\left[1-\cos\left(2\zeta\right)\right]\cos\left(2\xi_{F}\right)\right.\nonumber \\
 & +\left[4\zeta\sin\left(2\zeta\right)+5\cos\left(2\zeta\right)-5\right]\sin\left(2\xi_{F}\right)\nonumber \\
 & +4\left.\left[\zeta^{2}\left(\cos\left(2\zeta\right)+1\right)+g\left(\zeta\right)\right]\mathrm{si}\left(2\xi_{F}\right)\right\} ,\nonumber \\
 & F_{\mathrm{DM}}\left(\xi_{F},\zeta\right)\nonumber \\
= & \frac{-C}{\left(k_{F}R\right)^{4}}\left\{ 2\xi_{F}\sin\left(2\zeta\right)\cos\left(2\xi_{F}\right)\right.\nonumber \\
 & +\left[4\zeta\cos\left(2\zeta\right)-3\sin\left(2\zeta\right)\right]\sin\left(2\xi_{F}\right)\nonumber \\
 & -4\left.\left[\zeta^{2}\sin\left(2\zeta\right)+\zeta\left(\cos\left(2\zeta\right)-1\right)\right]\mathrm{si}\left(2\xi_{F}\right)\right\} ,\label{FHIDM}
\end{align}
where $g\left(x\right)=-2x\sin\left(2x\right)-\cos\left(2x\right)+1$,
$\xi_{F}=\sqrt{m^{2}\alpha^{2}+2m\varepsilon_{F}}R$, $k_{F}=\sqrt{2m\left|\varepsilon_{F}\right|}$
is the Fermi wave vector in the absence of SOC, $\mathrm{si}\left(z\right)=-\int_{z}^{\infty}\mathrm{d}t\sin t/t$
is the sine integral function, and $C=J^{2}mk_{F}^{4}/\left(2\pi\right)^{3}$.
The three range functions in Eq. $\left(\ref{FHIDM}\right)$ exhibit
several features. First, as shown in Fig.~$\ref{Fig2}$(a), for a
weak SOC, the Heisenberg interaction dominates over the other two
interactions within a moderately long range. Second, the magnitudes of Fermi wave
vectors of these two Fermi surfaces are unequal $k_{F-}-k_{F+}=2m\alpha$
for $\varepsilon_{F}>0$ or $k_{F+}-k_{F-}=2\sqrt{m^{2}\alpha^{2}+2m\varepsilon_{F}}$
for $\varepsilon_{F}<0$ such that the range functions in Eq.~$\left(\ref{FHIDM}\right)$
oscillate with two distinct periods and form beating patterns in Fig.~$\ref{Fig2}$(b). 
Third, while the range functions for $\varepsilon_{F}<0$ share similar features to those for $\varepsilon_{F}>0$,  
Figs.~$\ref{Fig2}$(c) and $\ref{Fig2}$(d) clearly show that the oscillation period quantitatively differs between the two cases. 
In addition, Fig.~$\ref{Fig2}$ shows that these range functions display a damped oscillatory behavior with increasing
distance $R$, with each term dominating in different regions of the
parameters $\varepsilon_{F}$, $R$, and $\alpha$. The competition
among these interactions leads to rich spin textures, which can be
detected by a variety of experimental tools. 

To gain more insight into the impacts of SOC on the RKKY interactions,
let us examine the weak SOC limit $\alpha k_{F}/\varepsilon_{F}\ll1$.
Expanding the analytical expressions of range functions in Eq.~$\left(\ref{FHIDM}\right)$
in power of $\alpha$ and keeping the correction up to $\alpha^{2}$,
one finally gets the following approximate range functions:  
\begin{align}
F_{\mathrm{H}} & \simeq4F_{0}\left(R\right)-4C\zeta^{2}h\left(k_{F}R\right),\nonumber \\
F_{\mathrm{Ising}} & \simeq8F_{0}\left(R\right)\zeta^{2},\nonumber \\
F_{\mathrm{DM}} & \simeq-8F_{0}\left(R\right)\zeta,\label{rfncapprox}
\end{align}
where $h\left(x\right)=\left[x\cos\left(2x\right)+\sin\left(2x\right)\right]/x^{4}$,
$F_{0}\left(R\right)$ is the range function for the conventional
3DEG \cite{Ruderman1954Kittel,Kasuya1956,Yosida1957}: 
\begin{equation}
F_{0}\left(R\right)=\frac{C\left[2k_{F}R\cos\left(2k_{F}R\right)-\sin\left(2k_{F}R\right)\right]}{4\left(k_{F}R\right)^{4}}.\label{RF3deg}
\end{equation}
According to Eq.~$\left(\ref{rfncapprox}\right)$, the Ising interaction
and DM interaction are quadratic and linear in $\zeta$, respectively.
Thus, in the weak SOC limit $\alpha k_{F}/\varepsilon_{F}\ll1$,
the Heisenberg interaction becomes dominant as that of the 3DEG. It qualitatively
coincides with the behavior of the exact range functions depicted
in Fig.~$\ref{Fig2}$(a). 
When the magnetic impurities are dilute, the RKKY interaction is mainly
controlled by its long-range behavior. Let us turn to the long-range
case of $\xi_{F}\gg1$ and $\xi_{F}\gg\zeta$, which are equivalent
to $R\gg1/\sqrt{m^{2}\alpha^{2}+2m\varepsilon_{F}}$ and $\sqrt{m^{2}\alpha^{2}+2m\varepsilon_{F}}\gg m\alpha$,
respectively. Under these conditions, we have $\mathrm{si}\left(2\xi_{F}\right)\simeq-\cos\left(2\xi_{F}\right)/2\xi_{F}$
and hence $\xi_{F}\left|\cos\left(2\xi_{F}\right)\right|\gg\zeta\left|\cos\left(2\xi_{F}\right)\right|\gg\zeta^{2}\left|\mathrm{si}\left(2\xi_{F}\right)\right|$.
As a result, the long-range RKKY interaction can be cast into a simple twisted form: 
\begin{equation}
H_{\mathrm{RKKY}}^{\mathrm{NCS}}\simeq\frac{2C\xi_{F}\cos\left(2\xi_{F}\right)}{\left(k_{F}R\right)^{4}}\boldsymbol{S}_{1}\cdot\tilde{\boldsymbol{S}}_{2},\label{eq:twistedrkky}
\end{equation}
where $\tilde{\boldsymbol{S}}_{2}$ stands for the twisted spin operator of the second localized magnetic
impurity and can be obtained from a rotation in spin space around
the $j$ axis (the direction of $\boldsymbol{R}$) by an angle $2\zeta$~
\cite{Aleiner2001PRL}. For example, taking $\boldsymbol{R}=R\boldsymbol{e}_{x}$,
after such a rotation the $x$ component of $\boldsymbol{S}_{2}$
remains unchanged, but the other two components become $\tilde{S}_{2}^{y}=\cos\left(2\zeta\right)S_{2}^{y}-\sin\left(2\zeta\right)S_{2}^{z}$
and $\tilde{S}_{2}^{z}=\sin\left(2\zeta\right)S_{2}^{y}+\cos\left(2\zeta\right)S_{2}^{z}$.
It is clear that the SOC tends to twist the spin to form a collinear coupling of localized spins. 
This kind of twisted exchange interaction was also discussed in the
context of 1DEG and 2DEG with standard Rashba SOC. However, their range functions essentially differ from ours. 
Specifically, the corresponding range functions in 1D and 2D are proportional to $\mathrm{si}(2\xi_{F})$ and $\sin(2\xi_{F})/R^2$, respectively~\cite{Imamura2004PRB}.  
In addition, the range function in Eq.~$\left(\ref{eq:twistedrkky}\right)$
is the same as that of 3DEG at large distance by replacing $k_{F}R$
with $\xi_{F}$, up to a degeneracy factor. Furthermore, it should
be noted that the spatial dependence of $R^{-3}$ is the same as both
3DEG~\cite{Aristov1997PRB} and 3D Dirac/Weyl semimetals~\cite{ChangHR2015PRB,Hosseini2015PRBWeyl,Mastrogiuseppe2016PRB}. 

\begin{figure}
\includegraphics[scale=0.65]{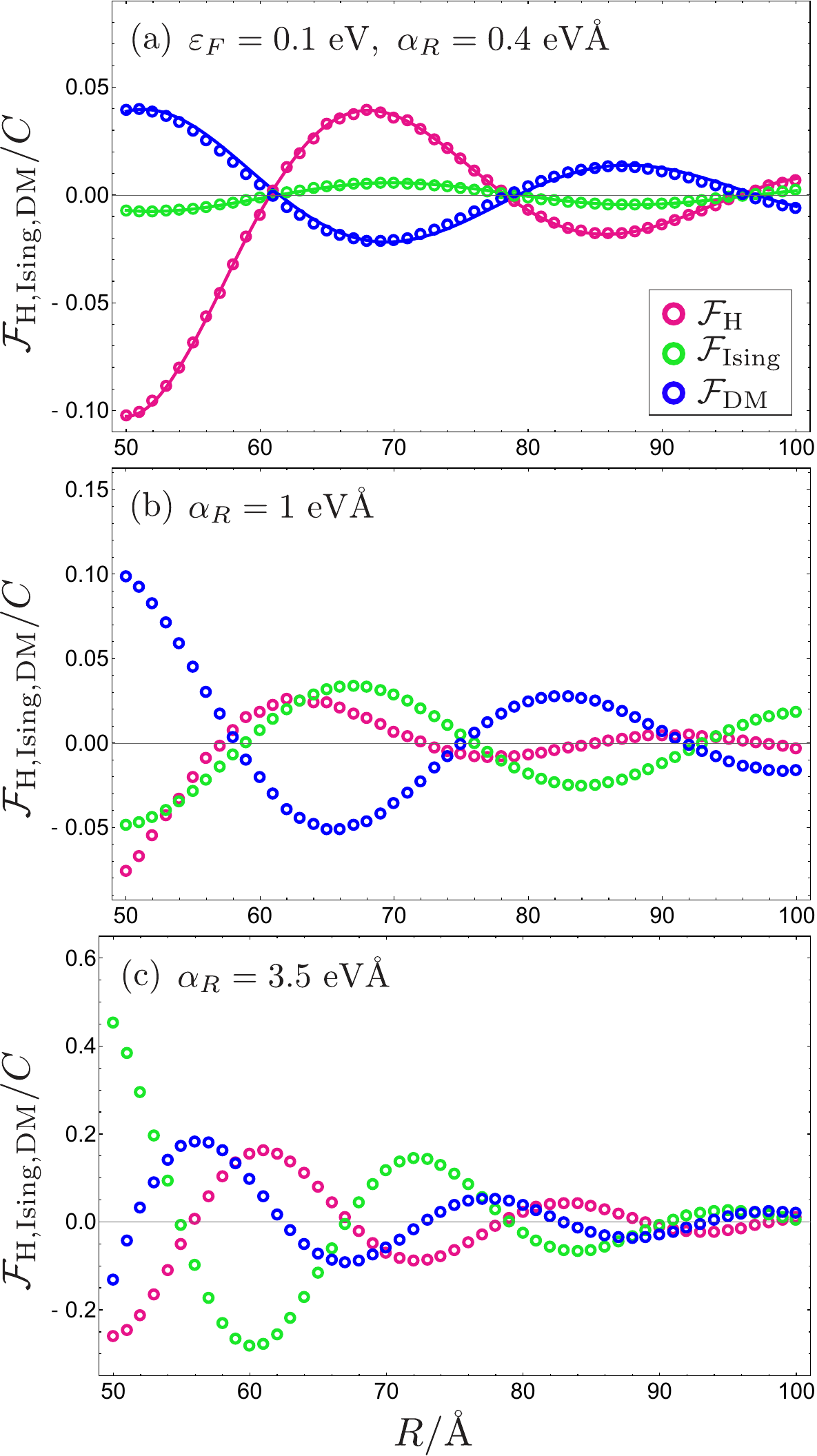}\caption{(Color online) RKKY interactions of 3D Rashba semiconductors BiTe$X$ for different
strengths of Rashba SOCs. (a) shows that the Heisenberg interaction dominates over the Ising and DM interactions within a large range. 
The solid lines in (a) correspond to the approximate range functions at zero temperature. 
For a strong Rashba SOC, these three interactions compete with each other in (b) and (c). 
Other parameters are given in Ref.~\cite{3dRSM}.
\label{Fig3}}
\end{figure}
%
\section{\label{sec:3drashba} three-dimensional rashba semiconductors}

The electrons in 3D Rashba semiconductors BiTe$X$ can be
described by the following effective Hamiltonian~\cite{Ishizaka2011BiTeI,Maiti2015PRB,bahramy2011PRB}: 
\begin{align}
H_{0}^{\mathrm{3DR}} & =\frac{k_{x}^{2}+k_{y}^{2}}{2m}+\frac{k_{z}^{2}}{2m_{3}}+\alpha_{R}\left(\boldsymbol{\sigma}\times\boldsymbol{k}\right)_{z},\label{eq:model2}
\end{align}
and the corresponding energy dispersion is given as 
\begin{equation}
\epsilon_{\lambda\boldsymbol{k}}=\frac{k_{x}^{2}+k_{y}^{2}}{2m}+\frac{k_{z}^{2}}{2m_{3}}+\lambda\alpha_{R}\sqrt{k_{x}^{2}+k_{y}^{2}},
\end{equation}
where the superscript $\mathrm{3DR}$ refers to the 3D Rashba semiconductors,
$\lambda=\pm$ stands for the chirality of energy bands, $\alpha_{R}$
is the strength of Rashba SOC, and $m$ and $m_{3}$ denote for the
effective masses in the $x$-$y$ plane and along the $z$ direction,
respectively. A series of experiments show that the ratio of $m_{3}/m$
ranges from $5$ to $10$. As shown in Fig.~$\ref{Fig1}$(c), for
$\varepsilon_{F}>0$, there are two distinct Fermi surfaces (spindle
torus), whereas only one closed Fermi surface (ring torus) exists
for $\varepsilon_{F}<0$~\cite{Landolt2012PRL}. This topological
transition of the bulk Fermi surface as a function of the Fermi energy
had been detected through Shubnikov\textendash de Haas quantum oscillation
\cite{Xiang2015PRB,YeLD2015PRB} and thermoelectric effects~\cite{Ideue2015PRB}.
The kinds of semiconductors described by Eq.~$\left(\ref{eq:model2}\right)$
can be intuitively seen as a coherent superposition of layers of 2DEG
with Rashba SOC along the $z$-direction. Unlike noncentrosymmetric
metals, the Rashba SOC in Eq.~$\left(\ref{eq:model2}\right)$ only
entangles $k_{x}$ and $k_{y}$ with $\sigma_{y}$ and $\sigma_{x}$,
respectively, but leaves $k_{z}$ and $\sigma_{z}$ free. Therefore,
the magnetism for in-plane magnetic impurities should significantly
differ from that for out-plane ones. 
In the bismuth tellurohalides family BiTe$X$ ($X$ = Br, Cl, I),
the structure inversion asymmetry results from the asymmetric stacking
of Bi, Te, and $X$ layers, in which an exceptionally large Rashba
parameter and $\pi$ Berry phase~\cite{XiaoRMPBerry2010}
had been revealed by photoemission studies~\cite{Ishizaka2011BiTeI}
and quantum oscillation experiment~\cite{murakawa2013Science}, respectively.
The strengths of Rashba SOC have been obtained experimentally: $3.9-4.3~\mathrm{eV\mathring{A}}$
for BiTeI, $2.0-2.1~\mathrm{eV\mathring{A}}$ for BiTeBr, and $1.7-2.2~\mathrm{eV\mathring{A}}$
for BiTeCl~\cite{Sakano2013PRL}. 

We shall consider the general case with the relative position vector
between magnetic impurities $\boldsymbol{R}=R_{x}\boldsymbol{e}_{x}+R_{y}\boldsymbol{e}_{y}+R_{z}\boldsymbol{e}_{z}$.
Following the similar procedure, one gets the RKKY interaction as
\begin{align}
H_{\mathrm{RKKY}}^{\mathrm{3DR}} & =\mathcal{F}_{\mathrm{H}}\boldsymbol{S}_{1}\cdot\boldsymbol{S}_{2}+\mathcal{F}_{\mathrm{\mathrm{Ising}}}\left(S_{1}^{x}\sin\varphi-S_{1}^{y}\cos\varphi\right)\nonumber \\
 & \times\left(S_{2}^{x}\sin\varphi-S_{2}^{y}\cos\varphi\right)+\mathcal{F}_{\mathrm{\mathrm{DM}}}\nonumber \\
 & \times\left[\left(\boldsymbol{S}_{1}\times\boldsymbol{S}_{2}\right)_{x}\sin\varphi-\left(\boldsymbol{S}_{1}\times\boldsymbol{S}_{2}\right)_{y}\cos\varphi\right],\label{rkky3dR}
\end{align}
where these range functions have the same form as those in Eq.~$\left(\ref{FhIdm}\right)$
but different real-space Green's function $G\left(\pm\boldsymbol{R};\varepsilon+i0^{+}\right)=G_{0}\left(R;\varepsilon+i0^{+}\right)\sigma_{0}\pm G_{R}\left(R;\varepsilon+i0^{+}\right)\left(\sigma_{x}\sin\varphi-\sigma_{y}\cos\varphi\right)$
with $\varphi=\arctan\left(R_{y}/R_{x}\right)$. In contrast to the
noncentrosymmetric metals, the DM vector in Eq.~$\left(\ref{rkky3dR}\right)$
lies in the $x$-$y$ plane and is perpendicular to the line connecting two magnetic impurities. 

\begin{figure}
\includegraphics[scale=0.5]{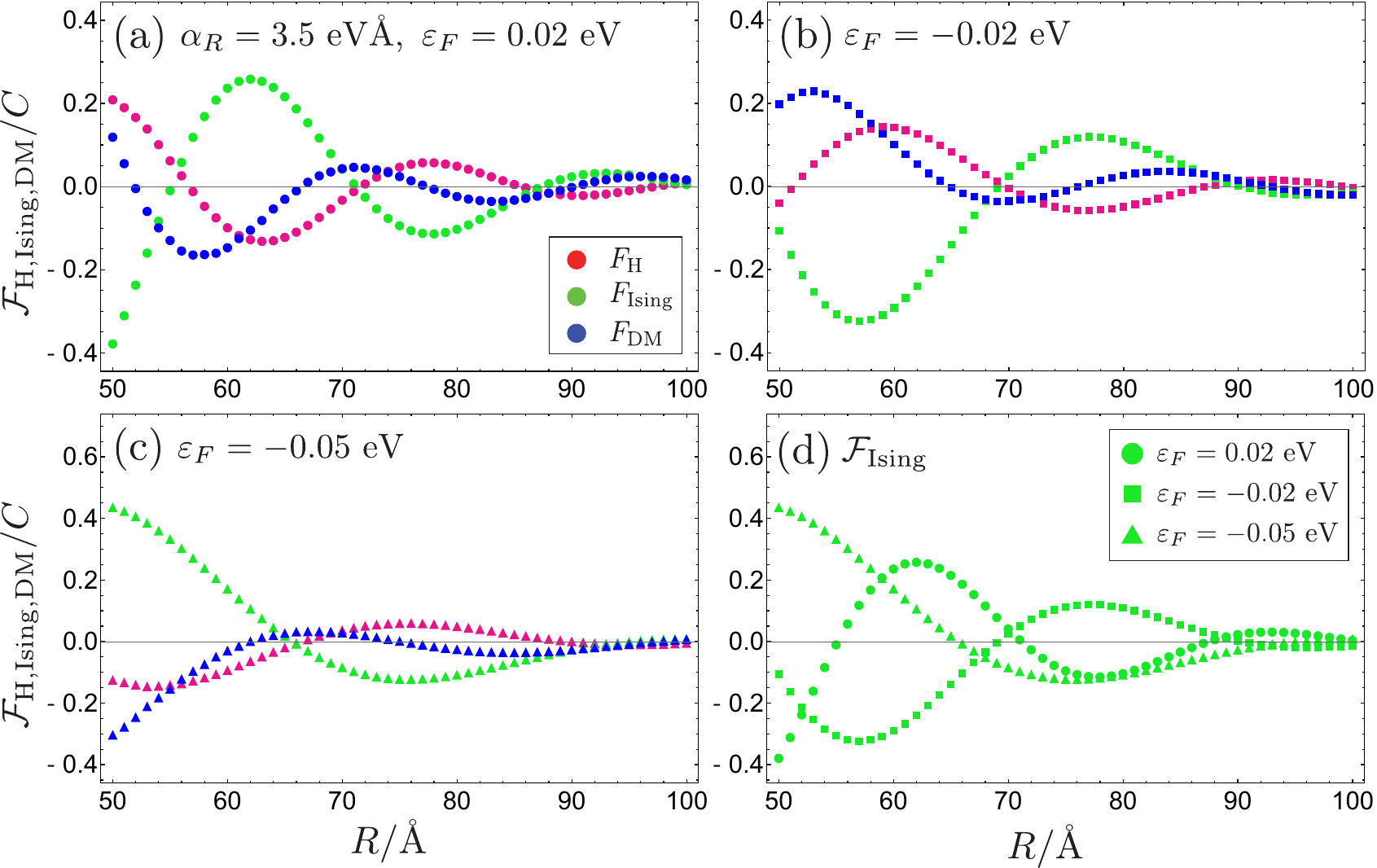}\caption{
(Color online) The evolution of RKKY interactions of 3D Rashba semiconductors BiTe$X$ for different values of Fermi energies.  
(a) and (c) show the suppression of the Heisenberg interaction for a small Fermi energy and a strong Rashba SOC. 
The dependence of the periods of oscillation of the Ising interaction on the Fermi energy  is depicted in (d). 
Other parameters are identical to those in Fig. \ref{Fig3}. \label{Fig4} }
\end{figure}

Direct numerical calculations of the range functions in Eq.~$\left(\ref{rkky3dR}\right)$
show several main features of the RKKY interactions of 3D Rashba semiconductors
\cite{3dRSM}. First, for a positive Fermi energy $\varepsilon_{F}>0$
and a weak Rashba SOC, the Heisenberg interaction dominates over the
other two interactions within a relatively large range in Fig.~\ref{Fig3}(a). 
Second, as the Rashba SOC increases, the Ising interaction and DM interaction would be enhanced gradually.
Eventually, these three terms compete with each other. In some regions,
the Ising interaction or the DM interaction dominates over the Heisenberg
interaction, as shown in Figs.~\ref{Fig3}(b) and \ref{Fig3}(c). Third, Figs.~\ref{Fig4}(a)-\ref{Fig4}(c) 
show that these three interactions also compete with each other as
the Fermi energy varies. For a small $\varepsilon_{F}$ and a strong Rashba
SOC, the Heisenberg interaction is greatly suppressed within a large
range of distance. 
In addition, Fig. \ref{Fig4}(d) depicts change in the periods of oscillation of the Ising interaction as the Fermi energy varies.
In other words, the magnetism of magnetic impurities
can be effectively manipulated through tuning the carrier concentration. 

Due to the great complexity of general expressions of the range functions,
in the present paper, we mainly focus on the analytical expressions
in the weak Rashba SOC case $\alpha_{R}k_{F}/\varepsilon_{F}\ll1$
and perturbatively treat the Rashba SOC. Keeping the contribution
of Rashba SOC up to $\alpha_{R}^{2}$, one arrives at the approximate
range functions (the detailed calculations are given in Appendix~\ref{sec:App3dRashba}): 
\begin{align}
\mathcal{F}_{\mathrm{H}} & \simeq4 \gamma^2 F_{0}\left(\tilde{R}\right)-8\gamma^2 m^{2}\alpha_{R}^{2}\nonumber \\
 & \times\left[\left(R_{x}^{2}+R_{y}^{2}\right)F_{0}\left(\tilde{R}\right)+\frac{C\sin\left(2k_{\mathrm{F}}\tilde{R}\right)}{2k_{\mathrm{F}}^{4}\tilde{R}^{2}}\right],\nonumber \\
\mathcal{F}_{\mathrm{\mathrm{Ising}}} & \simeq8 \gamma^2 m^{2}\alpha_{R}^{2}\left(R_{x}^{2}+R_{y}^{2}\right)F_{0}\left(\tilde{R}\right),\nonumber \\
\mathcal{F}_{\mathrm{\mathrm{DM}}} & \simeq-8 \gamma^2 m\alpha_{R}\sqrt{R_{x}^{2}+R_{y}^{2}}F_{0}\left(\tilde{R}\right),~\label{rf3dr}
\end{align}
where $\tilde{R}=\sqrt{R_{x}^{2}+R_{y}^{2}+\gamma^2 R_{z}^{2}}$, 
and $\gamma = \sqrt{m_{3}/m}$ measures the anisotropy of the effective mass.  
One immediately recognizes that the anisotropic effective mass may result in an anisotropic RKKY interaction.  
In addition, both sets of range functions in Eq.~$\left(\ref{rfncapprox}\right)$ and Eq.~$\left(\ref{rf3dr}\right)$
share a similar dependence on the strength of SOC. It should be
noted that the approximate range functions are consistent with the exact numerical results (hollow circles)
with a small Rashba SOC, as shown in Fig.~\ref{Fig3}(a). It implies that the low finite temperature only leads to 
small quantitative modifications. 

More interestingly, when the two magnetic impurities locate on a line
perpendicular to the $x$-$y$ plane ($\boldsymbol{R}=R_{z}\boldsymbol{e}_{z}$),
only the Heisenberg interaction survives and has the range function as
\begin{equation}
\mathcal{F}_{\mathrm{H}}\simeq4 \left[\gamma^2 F_{0}\left(\gamma R_{z}\right)-\frac{Cm^{2}\alpha_{R}^{2}\sin\left(2\gamma k_{F}R_{z}\right)}{k_{F}^{4}R_{z}^{2}}\right].\label{Ffm2dR}
\end{equation}
The vanishing of the Ising interaction and the DM interaction can
be traced back to the fact that the Rashba SOC only entangles the
in-plane spin components and in-plane momentum, as schematically shown
in Figs.~\ref{Fig1}(c) and \ref{Fig1}(d). Hence, the special form of Rashba SOC leads
to the intrinsic anisotropy of the RKKY interactions. One also finds
that the range function with an isotropic mass has a modification quadratic in $\alpha_{R}$
compared with $F_{0}\left(R_{z}\right)$ for the conventional
3DEG. When the range function is negative, the magnetic impurities
spontaneously align in parallel. Once the spin of magnetic impurities
is not perpendicular to the $z$ direction, the resulting magnetization
would give rise to a finite anomalous Hall current~\cite{Culcer2003PRB}. 

%
\section{\label{sec:conclus}conclusions}

In summary, we have demonstrated that two kinds of linear SOCs lead to the distinct RKKY interaction, 
including the Heisenberg interaction, the DM interaction, and the Ising interaction. 
In the weak SOC limit, the Heisenberg interaction 
is dominant over the other two interactions in a moderately large
region of parameters. The sufficiently strong SOC makes these
three interactions compete with each other. We have revealed that
the change in topology of the Fermi surface is accompanied by a shift
in periods of oscillations of range functions. The anisotropy of RKKY
interaction of the family BiTe$X$ comes from both the special Rashba
SOC and the anisotropy of the effective mass. Our work provides some
insights into understanding the nontrivial spin textures and paves
the way for the application of 3D noncentrosymmetric materials in spintronics. 

\section*{ACKNOWLEDGMENTS }

We thank Wen-Yu Shan for useful discussions. J.Z. was supported
by the Research Grant Council, University Grants Committee, Hong Kong
under Grants No. 17301116 and No. C6026-16W. S.-X.W. and H.-R.C.
were supported by the National Natural Science Foundation of China
under Grant No. 11547200. H.-R.C was also supported by the China
Scholarship Council, the NSERC of Canada, and FQRNT of Quebec (H. Guo).
We thank Compute Canada and the High Performance Computing Center
of McGill University where the numerical part was done.
%
\appendix

\section{RKKY interactions for 3D NCS metals \label{sec:AppNCS}}

In this Appendix we consider the RKKY interaction between two magnetic
impurities in NCS metals. The Green's function in momentum space is given by
\begin{align}
G\left(\boldsymbol{k};\varepsilon+i0^{+}\right) & =\frac{1}{2}\sum_{\lambda=\pm}\frac{\sigma_{0}+\lambda\hat{\boldsymbol{k}}\cdot\boldsymbol{\sigma}}{\varepsilon-\varepsilon_{\lambda\boldsymbol{k}}+i0^{+}},
\end{align}
where $\hat{\boldsymbol{k}}=\boldsymbol{k}/k$, and $\varepsilon_{\lambda\boldsymbol{k}}=\frac{k^{2}}{2m}+\lambda\alpha k$.
Thus, the Green's function in real space reads
\begin{align}
 & G\left(\pm\boldsymbol{R};\varepsilon+i0^{+}\right)\nonumber \\
= & \frac{1}{2}\sum_{\lambda=\pm}\int\frac{\mathrm{d}^{3}\boldsymbol{k}}{(2\pi)^{3}}\frac{\sigma_{0}\pm\lambda\hat{\boldsymbol{k}}\cdot\boldsymbol{\sigma}}{\varepsilon-\varepsilon_{\lambda\boldsymbol{k}}+i0^{+}}\exp\left(i\boldsymbol{k}\cdot\boldsymbol{R}\right).
\end{align}
In general, the momentum $\boldsymbol{k}$ can be decomposed as $\boldsymbol{k}=\boldsymbol{k}_{\parallel}+\boldsymbol{k}_{\perp}=\bigl(\hat{\boldsymbol{R}}\cdot\boldsymbol{k}\bigr)\hat{\boldsymbol{R}}+\bigl(\hat{\boldsymbol{R}}\times\boldsymbol{k}\bigr)\times\hat{\boldsymbol{R}}$,
where $\hat{\boldsymbol{R}}=\boldsymbol{R}/R$ is the direction of
$\boldsymbol{R}$ and $\boldsymbol{k}\cdot\boldsymbol{R}=kR\cos\theta$.
After integrating over $\phi$, one finds that the part of $\boldsymbol{k}_{\perp}$
vanishes and has
\begin{equation}
G\left(\pm\boldsymbol{R};\varepsilon+i0^{+}\right)=G_{0}\sigma_{0}\pm G_{R}\boldsymbol{\sigma}\cdot\hat{\boldsymbol{R}}\label{eq:decomp}
\end{equation}
where 
\begin{align}
G_{0} & =\frac{1}{2}\sum_{\lambda=\pm}\int\frac{k^{2}\sin\theta\mathrm{d}k\mathrm{d}\theta}{(2\pi)^{2}}\frac{\exp\left(ikR\cos\theta\right)}{\varepsilon-\varepsilon_{\lambda\boldsymbol{k}}+i0^{+}},\\
G_{R} & =\frac{1}{2}\sum_{\lambda=\pm}\int\frac{k^{2}\sin\theta\mathrm{d}k\mathrm{d}\theta}{(2\pi)^{2}}\frac{\lambda}{ik}\frac{\partial}{\partial R}\frac{\exp\left(ikR\cos\theta\right)}{\varepsilon-\varepsilon_{\lambda\boldsymbol{k}}+i0^{+}}.
\end{align}
Integrating over $\theta$ yields
\begin{align}
G_{0} & =-\frac{i}{2\left(2\pi\right)^{2}R}\sum_{\lambda=\pm}\int_{-\infty}^{\infty}\frac{k\exp\left(ikR\right)\mathrm{d}k}{\varepsilon-\varepsilon_{\lambda\boldsymbol{k}}+i0^{+}},\nonumber \\
G_{R} & =-\frac{1}{2\left(2\pi\right)^{2}}\frac{\partial}{\partial R}\left[\frac{1}{R}\sum_{\lambda=\pm}\lambda\int_{-\infty}^{\infty}\frac{\exp\left(ikR\right)\mathrm{d}k}{\varepsilon-\varepsilon_{\lambda\boldsymbol{k}}+i0^{+}}\right].\label{eq:gr-1}
\end{align}
Note that we have used the property of even function to extend the
range of integral to the whole real axis. In this way, all the remaining
calculations become standard contour integrals and can be carried out immediately:
\begin{align}
 & \int_{-\infty}^{\infty}\frac{k\exp\left(ikR\right)\mathrm{d}k}{\varepsilon-\varepsilon_{\lambda\boldsymbol{k}}+i0^{+}}\nonumber \\
= & -\int_{-\infty}^{\infty}\frac{2mk\exp\left(ikR\right)\mathrm{d}k}{\left(k-k_{-}^{\lambda}+i0^{+}\right)\left(k-k_{+}^{\lambda}-i0^{+}\right)}\nonumber \\
= & -\frac{2m\left(2\pi i\right)k_{+}^{\lambda}\exp\left(ik_{+}^{\lambda}R\right)}{k_{+}^{\lambda}-k_{-}^{\lambda}+i0^{+}}\nonumber \\
= & -i2\pi m\left(1-\frac{\lambda\zeta}{\xi+i0^{+}}\right)\exp\left(i\xi-i\lambda\zeta\right),
\end{align}
where $k_{\pm}^{\lambda}=\pm\sqrt{m^{2}\alpha^{2}+2m\varepsilon}-\lambda m\alpha$,
$\xi=\sqrt{m^{2}\alpha^{2}+2m\varepsilon}R$, and $\zeta=m\alpha R$.
Similarly, we have 
\begin{equation}
\int_{-\infty}^{\infty}\frac{\exp\left(ikR\right)\mathrm{d}k}{\varepsilon-\varepsilon_{\lambda\boldsymbol{k}}+i0^{+}}=-\frac{i2\pi mR}{\xi+i0^{+}}\exp\left(i\xi-i\lambda\zeta\right).
\end{equation}
Plugging these results into Eq.~$\left(\ref{eq:gr-1}\right)$, we find
\begin{align}
G_{0} & =-\frac{m\exp\left(i\xi\right)}{2\pi R\left(\xi+i0^{+}\right)}\left(\xi\cos\zeta+i\zeta\sin\zeta\right),\nonumber \\
G_{R} & =\frac{m\exp\left(i\xi\right)}{2\pi R\left(\xi+i0^{+}\right)}\left[\left(i\xi-1\right)\sin\zeta+\zeta\cos\zeta\right].\label{G0R}
\end{align}
Inserting the real-space Green's functions in Eq.~$\left(\ref{eq:decomp}\right)$
into Eq.~$\left(\ref{eq:rkky}\right)$, and using the trace formulas
for Pauli matrices $\mathrm{Tr}\left[\sigma_{i}\sigma_{j}\right]=2\delta_{ij}$,
$\mathrm{Tr}\left[\sigma_{i}\sigma_{j}\sigma_{l}\right]=2i\varepsilon_{ijl}$,
and $\mathrm{Tr}\left[\sigma_{i}\sigma_{l}\sigma_{j}\sigma_{m}\right]=2\left(\delta_{il}\delta_{jm}+\delta_{im}\delta_{jl}-\delta_{ij}\delta_{lm}\right)$,
we get
\begin{align}
H_{\mathrm{RKKY}}^{\mathrm{NCS}}= & F_{\mathrm{H}}\boldsymbol{S}_{1}\cdot\boldsymbol{S}_{2}+F_{\mathrm{Ising}}\left(\boldsymbol{S}_{1}\cdot\hat{\boldsymbol{R}}\right)\left(\boldsymbol{S}_{2}\cdot\hat{\boldsymbol{R}}\right)\nonumber \\
 & +F_{\mathrm{DM}}\left(\boldsymbol{S}_{1}\times\boldsymbol{S}_{2}\right)\cdot\hat{\boldsymbol{R}},
\end{align}
where the range functions are of the form
\begin{align}
 & F_{\mathrm{H}}\nonumber \\
= & -\frac{2J^{2}}{\pi}\mathrm{Im}\int_{-\infty}^{\varepsilon_{F}}\left(G_{0}^{2}+G_{R}^{2}\right)\mathrm{d}\varepsilon\nonumber \\
= & \frac{J^{2}m}{2\pi^{3}R^{4}}\Bigl\{\cos\left(2\zeta\right)\mathcal{I}_{2}\left(2\right)-\left[2\zeta\sin\left(2\zeta\right)+\cos\left(2\zeta\right)-1\right]\mathcal{I}_{1}\left(2\right)\nonumber \\
 & -\frac{1}{2}\left[2\zeta^{2}\cos\left(2\zeta\right)+g\left(\zeta\right)\right]\mathcal{I}_{0}\left(2\right)\Bigr\},\label{eq:f1r}\\
 & F_{\mathrm{Ising}}\nonumber \\
= & \frac{4J^{2}}{\pi}\mathrm{Im}\int_{-\infty}^{\varepsilon_{F}}G_{R}^{2}\mathrm{d}\varepsilon\nonumber \\
= & \frac{J^{2}m}{2\pi^{3}R^{4}}\Bigl\{2\left[\zeta\sin\left(2\zeta\right)+\cos\left(2\zeta\right)-1\right]\mathcal{I}_{1}\left(2\right)\nonumber \\
 & +\left[1-\cos\left(2\zeta\right)\right]\mathcal{I}_{2}\left(2\right)+\left[\zeta^{2}\left(\cos\left(2\zeta\right)+1\right)+g\left(\zeta\right)\right]\mathcal{I}_{0}\left(2\right)\Bigr\},\label{eq:f2r}
 \end{align}
 \begin{align}
 & F_{\mathrm{DM}}\nonumber \\
= & \frac{4J^{2}}{\pi}\mathrm{Im}\int_{-\infty}^{\varepsilon_{F}}iG_{0}G_{R}\mathrm{d}\varepsilon\nonumber \\
= & \frac{-J^{2}m}{2\pi^{3}R^{4}}\Bigl\{\sin\left(2\zeta\right)\mathcal{I}_{2}\left(2\right)+\left[2\zeta\cos\left(2\zeta\right)-\sin\left(2\zeta\right)\right]\mathcal{I}_{1}\left(2\right)\nonumber \\
 & -\left[\zeta^{2}\sin\left(2\zeta\right)+\zeta\left(\cos\left(2\zeta\right)-1\right)\right]\mathcal{I}_{0}\left(2\right)\Bigr\},\label{eq:f3r}
\end{align}
with $g\left(x\right)=-2x\sin\left(2x\right)-\cos\left(2x\right)+1$,
$\xi_{F}\equiv\sqrt{m^{2}\alpha^{2}+2m\varepsilon_{F}}R$. $\mathcal{I}_{0,1,2}\left(x\right)$
are given in Appendix \ref{sec:IntFormulas}. Substituting $\mathcal{I}_{0,1,2}\left(2\right)$
into Eqs.~$\left(\ref{eq:f1r}\right)$, $\left(\ref{eq:f2r}\right)$,
and $\left(\ref{eq:f3r}\right)$, we finally obtain the range functions in Eq.~(\ref{FHIDM}) in the main text. 
%
%
\section{approximate range functions for 3D Rashba semiconductors \label{sec:App3dRashba}}

In this Appendix, we calculate the RKKY interaction in 3D Rashba systems
up to the corrections due to Rashba SOC up to $\alpha_{R}^{2}$. The
momentum-space Green's function takes the form
\begin{equation}
G\left(\boldsymbol{k};\varepsilon+i0^{+}\right)=\frac{1}{2}\sum_{\lambda=\pm}\frac{\sigma_{0}+\lambda\left(\boldsymbol{\sigma}\times\hat{\boldsymbol{k}}_{\parallel}\right)_{z}}{\varepsilon-\epsilon_{\lambda\boldsymbol{k}}+i0^{+}},
\end{equation}
where $\boldsymbol{k}_{\parallel}=k_{x}\boldsymbol{e}_{x}+k_{y}\boldsymbol{e}_{y}$
denotes the in-plane component of $\boldsymbol{k}$, $\hat{\boldsymbol{k}}_{\parallel}=\boldsymbol{k}_{\parallel}/k_{\parallel}$,
$k_{\parallel}=\sqrt{k_{x}^{2}+k_{y}^{2}}$ and $\epsilon_{\lambda\boldsymbol{k}}=\frac{k^{2}}{2m}+\lambda\alpha_{R}k_{\parallel}$.
Here we adopt the isotropic mass $m_{3}=m$ at first. 

The real-space Green's function reads
\begin{align}
G\left(\pm\boldsymbol{R};\varepsilon+i0^{+}\right)= & \frac{1}{2}\sum_{\lambda=\pm}\int\frac{\mathrm{d}^{3}\boldsymbol{k}}{(2\pi)^{3}}\frac{\sigma_{0}\pm\lambda\left(\boldsymbol{\sigma}\times\hat{\boldsymbol{k}}_{\parallel}\right)_{z}}{\varepsilon-\epsilon_{\lambda\boldsymbol{k}}+i0^{+}}\nonumber \\
 & \times\exp\left[i\left(\boldsymbol{k}_{\parallel}\cdot\boldsymbol{R}_{\parallel}+k_{z}R_{z}\right)\right],
\end{align}
where $\boldsymbol{R}=R_{x}\boldsymbol{e}_{x}+R_{y}\boldsymbol{e}_{y}+R_{z}\boldsymbol{e}_{z}$
and $\boldsymbol{R}_{\parallel}=R_{x}\boldsymbol{e}_{x}+R_{y}\boldsymbol{e}_{y}$
is the in-plane component of $\boldsymbol{R}$. Let us express the
in-plane component of $\boldsymbol{k}$ as $\boldsymbol{k}_{\parallel}=\left(k_{\parallel}\cos\phi\right)\hat{\boldsymbol{R}}_{\parallel}+\left(k_{\parallel}\sin\phi\right)\boldsymbol{e}_{z}\times\hat{\boldsymbol{R}}_{\parallel}$,
where $\hat{\boldsymbol{R}}_{\parallel}=\boldsymbol{R}_{\parallel}/\sqrt{R_{x}^{2}+R_{y}^{2}}$
is the in-plane direction of $\boldsymbol{R}$, $\phi$ is the relative
angle between $\boldsymbol{k}_{\parallel}$ and $\boldsymbol{R}_{\parallel}$.
After integrating over $\phi$, one finds that the second part of
$\boldsymbol{k}_{\parallel}$ vanishes and has the real-space Green's function
\begin{equation}
G\left(\pm\boldsymbol{R};\varepsilon+i0^{+}\right)=G_{0}\sigma_{0}\pm G_{R}\left(\boldsymbol{\sigma}\times\hat{\boldsymbol{R}}_{\parallel}\right)_{z},\label{AG3dR}
\end{equation}
with
\begin{align}
G_{0}= & \frac{1}{2}\sum_{\lambda=\pm}\int\frac{k^{2}\sin\theta\mathrm{d}k\mathrm{d}\theta}{\left(2\pi\right)^{2}}\frac{J_{0}\left(kR_{\parallel}\sin\theta\right)}{\varepsilon-\epsilon_{\lambda\boldsymbol{k}}+i0^{+}}\nonumber \\
 & \times\exp\left(ikR_{z}\cos\theta\right),\nonumber \\
G_{R}= & -\frac{i}{2}\frac{\partial}{\partial R_{\parallel}}\left[\sum_{\lambda=\pm}\lambda\int\frac{k\mathrm{d}k\mathrm{d}\theta}{\left(2\pi\right)^{2}}\frac{J_{0}\left(kR_{\parallel}\sin\theta\right)}{\varepsilon-\epsilon_{\lambda\boldsymbol{k}}+i0^{+}}\right.\nonumber \\
 & \left.\times\exp\left(ikR_{z}\cos\theta\right)\right],
\end{align}
where $J_{0}\left(z\right)$ is the zeroth-order Bessel function of the first kind, $k_{\parallel}=k\sin\theta$, and $k_{z}=k\cos\theta$. 

Due to the complexity of the integral, we expand the Green's
function as a power series of the Rashba SOC $\alpha_{R}$. The zeroth-order
term of $G_{0}$ takes the form
\begin{equation}
G_{00}=\int\frac{k^{2}\sin\theta\mathrm{d}k\mathrm{d}\theta}{\left(2\pi\right)^{2}}\frac{J_{0}\left(kR_{\parallel}\sin\theta\right)\exp\left(ikR_{z}\cos\theta\right)}{\varepsilon-\epsilon_{k}+i0^{+}},\label{eq:g00d}
\end{equation}
where $\epsilon_{k}=\frac{k^{2}}{2m}$. The first-order term
\begin{align}
G_{01} & =\frac{1}{2}\sum_{\lambda=\pm}\int\left(\lambda\alpha_{R}k\sin\theta\right)\frac{k^{2}\sin\theta\mathrm{d}k\mathrm{d}\theta}{\left(2\pi\right)^{2}}\nonumber \\
 & \times\frac{J_{0}\left(kR_{\parallel}\sin\theta\right)\exp\left(ikR_{z}\cos\theta\right)}{\left(\varepsilon-\epsilon_{k}+i0^{+}\right)^{2}}
\end{align}
vanishes after summing over band index $\lambda$, while the second-order term 
\begin{align}
G_{02} & =\int\left(\alpha_{R}k\sin\theta\right)^{2}\frac{k^{2}\sin\theta\mathrm{d}k\mathrm{d}\theta}{\left(2\pi\right)^{2}}\nonumber \\
 & \times\frac{J_{0}\left(kR_{\parallel}\sin\theta\right)\exp\left(ikR_{z}\cos\theta\right)}{\left(\varepsilon-\epsilon_{k}+i0^{+}\right)^{3}}
\end{align}
is nonvanishing. One can immediately conclude that the contributions
from upper and lower bands vanish all the odd-order terms of $G_{0}$.
Up to the leading corrections due to Rashba SOC, we then approximate
$G_{0}\simeq G_{00}+G_{02}$. On the contrary, we find that $G_{R}$
only contains odd-order terms. Therefore, the first nonvanishing
term of $G_{R}$ is linear in $\alpha_{R}$
\begin{align}
G_{R1} & =-i\frac{\partial}{\partial R_{\parallel}}\int\left(\alpha_{R}k\sin\theta\right)\frac{k\mathrm{d}k\mathrm{d}\theta}{\left(2\pi\right)^{2}}\nonumber \\
 & \times\frac{J_{0}\left(kR_{\parallel}\sin\theta\right)\exp\left(ikR_{z}\cos\theta\right)}{\left(\varepsilon-\epsilon_{k}+i0^{+}\right)^{2}}.
\end{align}
Note that we approximate $G_{R}$ as $G_{R}\simeq G_{R1}$. We then
calculate $G_{00}$, $G_{02}$, and $G_{R1}$ in a similar way. By
use of Eq.~$\left(\ref{eq:J0}\right)$, integrating over $\theta$ leads to
\begin{align}
G_{00} & =-\frac{i}{\left(2\pi\right)^{2}R}\int_{-\infty}^{\infty}\frac{k\exp\left(ikR\right)\mathrm{d}k}{\varepsilon-\epsilon_{k}+i0^{+}},
\end{align}
where $R=\sqrt{R_{\parallel}^{2}+R_{z}^{2}}$ and the last step is
similar to Eq.~$\left(\ref{eq:gr-1}\right)$. Carrying out the straightforward contour integral, we have 
\begin{equation}
G_{00}=-\frac{m\exp\left(ik_{\varepsilon}R\right)}{2\pi R},\label{eq:g00}
\end{equation}
where $k_{\varepsilon}=\sqrt{2m\varepsilon}$. For $G_{R1}$, the
integration over $\theta$ is the same as that of $G_{00}$ in Eq.~$\left(\ref{eq:g00d}\right)$. Hence one immediately gets
\begin{align}
G_{R1} & =-\frac{\alpha_{R}}{\left(2\pi\right)^{2}}\frac{\partial}{\partial R_{\parallel}}\left[\frac{1}{R}\int_{-\infty}^{\infty}\frac{k\exp\left(ikR\right)\mathrm{d}k}{\left(\varepsilon-\epsilon_{k}+i0^{+}\right)^{2}}\right].
\end{align}
Performing the standard contour integral, we arrive at the result
\begin{equation}
G_{R1}=\frac{imR_{\parallel}\exp\left(ik_{\varepsilon}R\right)}{2\pi R^{2}}\zeta^{\prime},\label{eq:gr1}
\end{equation}
where the dimensionless parameter $\zeta^{\prime}=m\alpha_{R}R$.
The last one, $G_{02}$, can be calculated similarly. Using Eq.~$\left(\ref{eq:J1}\right)$,
we integrate over $\theta$ and get
\begin{align}
G_{02} & =-\frac{i\alpha_{R}^{2}}{\left(2\pi\right)^{2}}\left\{ \frac{1}{R}\int_{-\infty}^{\infty}\frac{k^{3}\exp\left(ikR\right)\mathrm{d}k}{\left(\varepsilon-\epsilon_{k}+i0^{+}\right)^{3}}\right.\nonumber \\
 & \left.+\frac{\partial^{2}}{\partial R_{z}^{2}}\left[\frac{1}{R}\int_{-\infty}^{\infty}\frac{k\exp\left(ikR\right)\mathrm{d}k}{\left(\varepsilon-\epsilon_{k}+i0^{+}\right)^{3}}\right]\right\} .
\end{align}
After some algebra, we have 
\begin{equation}
G_{02}=-\frac{m\exp\left(ik_{\varepsilon}R\right)}{2\pi R}\left(\frac{i}{k_{\varepsilon}R+i0^{+}}-\frac{R_{\parallel}^{2}}{2R^{2}}\right)\zeta^{\prime2}.\label{eq:g02}
\end{equation}
Collecting the results in Eqs.~$\left(\ref{eq:g00}\right)$, $\left(\ref{eq:gr1}\right)$,
and $\left(\ref{eq:g02}\right)$, we have the following approximations: 
\begin{align}
G_{0} & \simeq-\frac{m\exp\left(ik_{\varepsilon}R\right)}{2\pi R}\left[1+\left(\frac{i}{k_{\varepsilon}R+i0^{+}}-\frac{R_{\parallel}^{2}}{2R^{2}}\right)\zeta^{\prime2}\right],\nonumber \\
G_{R} & \simeq\frac{imR_{\parallel}\exp\left(ik_{\varepsilon}R\right)}{2\pi R^{2}}\zeta^{\prime}.\label{eq:gr-2}
\end{align}
Since the real-space Green's function in Eq.~($\ref{AG3dR}$) has the same structure as that
in Eq.~$\left(\ref{eq:decomp}\right)$, the RKKY interaction in 3D Rashba systems takes a similar form:
\begin{align}
 & H_{\mathrm{RKKY}}^{\mathrm{3DR}}\nonumber \\
= & \mathcal{F}_{\mathrm{H}}\boldsymbol{S}_{1}\cdot\boldsymbol{S}_{2}+\mathcal{F}_{\mathrm{DM}}\left[\left(\boldsymbol{S}_{1}\times\boldsymbol{S}_{2}\right)_{x}\sin\varphi-\left(\boldsymbol{S}_{1}\times\boldsymbol{S}_{2}\right)_{y}\cos\varphi\right]\nonumber \\
 & +\mathcal{F}_{\mathrm{Ising}}\left(S_{1}^{x}\sin\varphi-S_{1}^{y}\cos\varphi\right)\left(S_{2}^{x}\sin\varphi-S_{2}^{y}\cos\varphi\right),
\end{align}
where $\varphi=\arctan\left(R_{y}/R_{x}\right)$ and the ranges functions
are given as
\begin{align}
\mathcal{F}_{\mathrm{H}} & =-\frac{2J^{2}}{\pi}\mathrm{Im}\int_{-\infty}^{\varepsilon_{F}}\left(G_{0}^{2}+G_{R}^{2}\right)\mathrm{d}\varepsilon,\nonumber \\
\mathcal{F}_{\mathrm{Ising}} & =\frac{4J^{2}}{\pi}\mathrm{Im}\int_{-\infty}^{\varepsilon_{F}}G_{R}^{2}\mathrm{d}\varepsilon,\nonumber \\
\mathcal{F}_{\mathrm{DM}} & =\frac{4J^{2}}{\pi}\mathrm{Im}\int_{-\infty}^{\varepsilon_{F}}iG_{0}G_{R}\mathrm{d}\varepsilon.\label{eq:fhidm-2}
\end{align}
Plugging Eq.~$\left(\ref{eq:gr-2}\right)$ into Eq.~$\left(\ref{eq:fhidm-2}\right)$
and using the basic integrals in Eqs.~$\left(\ref{eq:I1x}\right)$
and $\left(\ref{eq:I2x}\right)$, we obtain the following approximations
up to the leading corrections from SOC:
\begin{align}
\mathcal{F}_{\mathrm{H}} & \simeq4F_{0}\left(R\right)-8m^{2}\alpha_{R}^{2}\left[R_{\parallel}^{2}F_{0}\left(R\right)+\frac{C\sin\left(2k_{F}R\right)}{2k_{F}^{4}R^{2}}\right],\nonumber \\
\mathcal{F}_{\mathrm{Ising}} & \simeq8m^{2}\alpha_{R}^{2}R_{\parallel}^{2}F_{0}\left(R\right),\nonumber \\
\mathcal{F}_{\mathrm{DM}} & \simeq-8m\alpha_{R}R_{\parallel}F_{0}\left(R\right),
\end{align}
where $F_{0}\left(R\right)$ is the range function for 3DEG in Eq.~$\left(\ref{RF3deg}\right)$. 

The impacts of anisotropy of the effective mass $\gamma=\sqrt{m_{3}/m}\neq1$
in Eq.~$\left(\ref{eq:model2}\right)$ can be encoded into the RKKY
interaction through a transformation introduced in Ref.~\cite{ChangHR2015PRB}.
Specifically, the real-space Green's functions for the isotropic Hamiltonian
and the anisotropic one can be connected by the following transformation
$\left(k_{z},R_{z}\right)\rightarrow\left(\gamma k_{z},R_{z}/\gamma\right)$.
We note that the above transformation preserves the volume of the
Fermi sphere. After some algebra, one gets the relation $\tilde{G}\left(R_{x},R_{y},R_{z}\right)=\gamma G\left(R_{x},R_{y},\gamma R_{z}\right)$,
leading to the anisotropic RKKY interaction 
\begin{equation}
\tilde{H}_{\mathrm{RKKY}}\left(R_{x},R_{y},R_{z}\right)=\gamma^{2}H_{\mathrm{RKKY}}\left(R_{x},R_{y},\gamma R_{z}\right),
\end{equation}
which implies that the anisotropic effective mass may result in an anisotropic RKKY interaction. 
Thus, it is easy to reach the approximate range functions in Eq.~(\ref{rf3dr}). 

%
\section{some integral formulas \label{sec:IntFormulas}}

In this Appendix we evaluate two sets of frequently used integrals
$\mathcal{I}_{n}\left(x\right)$ and $\mathcal{J}_{n}$: 
\begin{align}
\mathcal{I}_{n}\left(x\right) & =\mathrm{Im}\int_{-\infty}^{\xi_{F}}\frac{i^{n}\exp\left(ix\xi\right)}{\left(\xi+i0^{+}\right)^{1-n}}\mathrm{d}\xi,\\
\mathcal{J}_{n} & =\int_{0}^{\pi}\left(\sin\theta\right)^{2n+1}J_{0}\left(kR_{\parallel}\sin\theta\right)\nonumber \\
 & \times\exp\left(ikR_{z}\cos\theta\right)\mathrm{d}\theta,
\end{align}
with $n$ being a nonnegative integer. For $n>1$, $\mathcal{I}_{n}\left(x\right)$
and $\mathcal{J}_{n}$ can be generated by $\mathcal{I}_{0}\left(x\right)$
and $\mathcal{J}_{0}$, respectively.
\begin{align}
\mathcal{I}_{n}\left(x\right) & =\frac{\partial^{n}}{\partial x^{n}}\left[\mathrm{Im}\int_{-\infty}^{\xi_{F}}\frac{\exp\left(ix\xi\right)}{\xi+i0^{+}}\mathrm{d}\xi\right]=\frac{\partial^{n}\mathcal{I}_{0}\left(x\right)}{\partial x^{n}},\label{eq:Inx}\\
\mathcal{J}_{n} & =\sum_{m=0}^{n}\frac{C_{n}^{m}}{k^{2m}}\frac{\partial^{2m}\mathcal{J}_{0}}{\partial R_{z}^{2m}},\label{eq:Jn}
\end{align}
where $C_{n}^{m}=\frac{n!}{m!\left(n-m\right)!}$ is the binomial
coefficient. Note that the infinitesimal $0^{+}$ in the numerator
in Eq.~$\left(\ref{eq:Inx}\right)$ can be safely omitted. Therefore,
our task is to compute $\mathcal{I}_{0}\left(x\right)$ and $\mathcal{J}_{0}$. 

Let us first deal with $\mathcal{I}_{0}\left(x\right)$ as follows: 
\begin{align}
 & \mathcal{I}_{0}\left(x\right)\nonumber \\
= & \mathrm{Im}\int_{-\infty}^{\xi_{F}}\frac{\exp\left(ix\xi\right)}{\xi+i0^{+}}\mathrm{d}\xi\nonumber \\
= & \mathrm{Im}\left[\mathcal{P}\int_{-\infty}^{\xi_{F}}\frac{\exp\left(ix\xi\right)}{\xi}\mathrm{d}\xi-i\pi\int_{-\infty}^{\xi_{F}}\exp\left(ix\xi\right)\delta\left(\xi\right)\mathrm{d}\xi\right]\nonumber \\
= & \frac{\pi}{2}\mathrm{sgn}\left(x\right)+\mathrm{Si}\left(x\xi_{F}\right)-\pi,
\end{align}
where we have used the identity $\frac{1}{x+i0^{+}}=\mathcal{P}(\frac{1}{x})-i\pi\delta\left(x\right),$
with $\mathcal{P}$ denoting the Cauchy principle value. $\mathrm{Si}\left(z\right)=\int_{0}^{z}\mathrm{d}t\sin t/t$
is the sine integral~\cite{Gradshteyn2007table}. Since only the case
$x>0$ is of interest, we have 
\begin{align}
\mathcal{I}_{0}\left(x\right) & =\frac{\pi}{2}+\mathrm{Si}\left(x\xi_{F}\right)-\pi=\mathrm{si}\left(x\xi_{F}\right),\label{eq:I0x}
\end{align}
with $\mathrm{si}\left(z\right)=-\int_{z}^{\infty}\mathrm{d}t\sin t/t=\mathrm{Si}\left(z\right)-\pi/2$.
Next, we turn to evaluate $\mathcal{J}_{0}$: 
\begin{align}
\mathcal{J}_{0} & =\int_{0}^{\pi}\sin\theta J_{0}\left(kR_{\parallel}\sin\theta\right)\exp\left(ikR_{z}\cos\theta\right)\mathrm{d}\theta\nonumber \\
 & =-\int_{0}^{\pi}J_{0}\left(kR_{\parallel}\sqrt{1-\cos^{2}\theta}\right)\exp\left(ikR_{z}\cos\theta\right)\mathrm{d}\left(\cos\theta\right)\nonumber \\
 & =\int_{-1}^{1}J_{0}\left(kR_{\parallel}\sqrt{1-t^{2}}\right)\exp\left(ikR_{z}t\right)\mathrm{d}t\nonumber \\
 & =2\int_{0}^{1}J_{0}\left(kR_{\parallel}\sqrt{1-t^{2}}\right)\cos\left(kR_{z}t\right)\mathrm{d}t.
\end{align}
Using the formula in Ref.~\cite{Gradshteyn2007table}, we get
\begin{equation}
\mathcal{J}_{0}=\frac{2\sin\left(kR\right)}{kR},\label{eq:J0}
\end{equation}
where $R=\sqrt{R_{\parallel}^{2}+R_{z}^{2}}$. From Eqs.~$\left(\ref{eq:Inx}\right)$
and $\left(\ref{eq:Jn}\right)$, we have
\begin{align}
\mathcal{I}_{1}\left(x\right) & =\frac{\partial\mathrm{si}\left(x\xi_{F}\right)}{\partial x}=\frac{\sin\left(x\xi_{F}\right)}{x},\label{eq:I1x}\\
\mathcal{I}_{2}\left(x\right) & =\frac{\partial^{2}\mathrm{si}\left(x\xi_{F}\right)}{\partial x^{2}}=\frac{\xi_{F}\cos\left(x\xi_{F}\right)}{x}-\frac{\sin\left(x\xi_{F}\right)}{x^{2}},\label{eq:I2x}
\end{align}
and 
\begin{equation}
\mathcal{J}_{1}=\left(1+\frac{1}{k^{2}}\frac{\partial^{2}}{\partial R_{z}^{2}}\right)\frac{2\sin\left(kR\right)}{kR},\label{eq:J1}
\end{equation}
which are used in Appendices A and B. Those who are interested in higher-order
corrections from SOC need to continue to look for the expressions of larger $n$.
%
\bibliographystyle{apsrev4-1}
%

%

\end{document}